\documentclass[12pt]{article}
\usepackage{amssymb,amsmath,epsfig,mathtools,caption}
\allowdisplaybreaks
\begin{document}

\title{\bf $f(T)$ Corrected Instability of Cylindrical Collapsing Object with Harrison-Wheeler Equation of State}
\author{Abdul Jawad$^1$ \thanks{jawadab181@yahoo.com; abduljawad@ciitlahore.edu.pk}
and M. Azam$^{2}$ \thanks{azam.math@ue.edu.pk} \\
\small $^1$Department of Mathematics, COMSATS Institute of Information\\
\small Technology, Lahore-54000, Pakistan.\\
\small $^2$Division of Science and Technology, University of
Education,\\ \small Township Campus, Lahore-54590, Pakistan}

\date{}
\maketitle

\begin{abstract}
In this paper, we study the dynamical instability of a collapsing
object in the framework of generalized teleparallel gravity. We
assume a cylindrical object with a specific matter distribution.
This distribution contains energy density, isotropic pressure
component with heat conduction. We take oscillating states scheme up
to first order to check the instable behavior of the object. We
construct a general collapse equation for underlying case with
non-diagonal tetrad depending on the matter, metric functions, heat
conducting term and torsional terms. The Harrison-Wheeler equation
of state which contains adiabatic index is used to explore the
dynamical instability ranges for Newtonian and post-Newtonian
constraints. These ranges depend on perturbed part of metric
coefficients, matter parts and torsion.
\end{abstract}

\section{Introduction}

General relativity is one of the most acceptable theory of gravity
which describes many natural phenomena of universe. However, this
theory faces some issues such as dark matter and dark energy. On the
same side, dark matter and dark energy are the important ingredients
for the dynamics of the entire universe. That is, 25\% part of
universe consists of dark matter while dark energy is upto 68\%. On
the other hand, dark energy is responsible for the expansion of the
universe. These problems are the main reason behind the modification
of general relativity (GR). The modification of GR is as old as this
theory itself. In recent era, $f(R)$ theory of gravity is the
simplest modification to GR based on curvature tensor. However, this
theory contains fourth order field equations which are very
difficult to handle.

An alternative theory to GR is teleparallel gravity which based on
torsion using Weitzenb\"{o}ck connection where curvature is zero.
The modification of telaparallel gravity results a generalized
Teleparallel gravity, that is $ f(T)$ theory of gravity
\cite{1}-\cite{8}. One of the advantage of this theory is that it
gives field equations of second order which are easier to handle for
the discussion of any underlying scenario as compared to $f(R)$
gravity. In this context, Ferraro and Fiorini introduced the $f(T)$
theory of gravity to solve the partial horizon problem in
Born-Infeld strategy which gave singularity free solution \cite{6}.
It is also known as non-local Lorentz invariant theory. o figure out
this problem, Nashed \cite{2} has used two tetrad matrices for the
regularization of $f(T)$ field equations. He proposed regularized
process with general tetrad field to figure out the effect of local
Lorentz invariance. Some authors also investigated this problem for
reference see \cite{7}-\cite{8}.

The dynamical collapse of self gravitating spherically symmetric
object has been widely discussed in $f(T)$ theory of gravity
\cite{9}-\cite{14}. The collapse process occurs when the balanced
matter of object become imbalanced and the object fails to maintain
its equilibrium. In this way, various dynamical states occur, which
may be analyzed through dynamical equations. Chandrasehkar \cite{15}
was the first who introduced the concept of dynamical instability
analysis with the help of adiabatic index, $\Gamma$ . Instability
ranges through adiabatic index have been critically examined for
cylindrically and spherically symmetric collapsing matter in $f(R)$
gravity \cite{16}-\cite{22}. Skripkin \cite{23} analyzed the
collapsing non-dissipative spherically symmetric fluid with constant
energy density and isotropy. He concluded that under expansion-free
condition, Minkowskian cavity is located at the center of the fluid.
Similarly under the same conditions, dynamical collapse analysis of
spherically and cylindrically symmetric anisotropic fluid is
explored by Chan et al. \cite{24}-\cite{27}. Some authors
\cite{28}-\cite{29} also studied the instability ranges of
spherically symmetric collapsing star with the presence of charge
and without charge in $f(R)$ gravity and concluded that these
instability ranges are based on geometry, matter and curvature.

In this scenario, Kausar worked on the effects of CDTT model having
inverse curvature term on the unpredictable behavior for the
cylindrical symmetric object. It was shown that dynamical
instability of expansion free gravitational collapse with spherical
geometry can be investigated without adiabatic index $\Gamma$
\cite{30}-\cite{31}. It was concluded that instability ranges depend
on energy density, electromagnetic field and anisotropic pressure.
In Brans-Dicke gravity, Sharif and Manzoor \cite{32} explore the
dynamical instability of cylindrically symmetric collapsing star and
concluded the range of adiabatic index which is greater than one in
special case and remains less than one for unstable behavior. In
$f(T)$ gravity, dynamics of collapsing spherical symmetric object
with expansion and expansion-free, shear-free conditions have been
explored \cite{13}-\cite{14}.

Jawad et al. \cite{13} discussed the dynamical instability ranges of
a cylindrical symmetric object in the framework of $f(T)$ gravity.
They have used anisotropic matter distribution and concluded that
the instability ranges effected by the presence of matter, metric
coefficients and torsional terms. We extend this paper taking matter
contribution which contain isotropic pressure with heat conducting
term. The scheme of the paper is as follows. In section \textbf{2},
we construct the field equations for cylindrical symmetric
collapsing object for $f(T)$ gravity taking non-diagonal tetrad.
Section \textbf{3} provides the perturbation scheme  up to first
order to develop a general collapse equation. In the next section
\textbf{4}, we obtain instability ranges under Newtonian and post
Newtonian constraints. The last section contains the concluding
remarks.

\section{Field Equations of $f(T)$ Gravity for Collapsing Star}

In this section, we provide the basic formulation of $f(T)$ gravity.
We discuss the basics of collapsing star in cylindrical symmetry and
obtain the $f(T)$ field equations in this scenario. We have taken
cylindrically symmetric line element as interior spacetime while
exterior spacetime in retarded time coordinate in the framework of
$f(T)$ gravity. The collapse process happens when stability of
matter disturbed and at long last experiences collapse which leads
to different structures. We have taken the self-gravitating object
as cylindrically symmetric collapsing star. We consider
cylindrically symmetric collapsing star as an interior region
defined by
\begin{equation}\label{31}
{{ds^{2}_{(-)}}}=-{\mathcal{A}^{2}dt^{2}}+{{\mathcal{\mathcal{B}}^{2}dr^{2}}}
+{{\mathcal{C}^{2}d{\phi}^{2}}}+{{dz^{2}}},
\end{equation}
where $\mathcal{A}, \mathcal{B}$ and $\mathcal{C}$ are functions of
$t,~r$ and cylindrical coordinates satisfy the following constraints
$\nonumber {-\infty}~{\leq}~{t}~{\leq}~{\infty},$ $\nonumber
{0}~{\leq}~{r}~{<}~{\infty},$ $\nonumber
{-\infty}~{<}~{z}~{<}~{\infty},
~~{0}~{\leq}~{\phi}~{\leq}~{2{\pi}}.$ In order to keep the
cylindrical symmetry, it must satisfy some conditions in the
beginning of collapse which is not a trivial process. That is, if
the symmetry does not contain any curvature singularity then we know
the conditions to apply. Otherwise, on the singular symmetry, it is
hard to impose any condition. In the underlying case, the inside
region of cylinder is assumed to be flat, that is axis is regular.
Thus, we may apply the following conditions \cite{ec}-\cite{ec1}.
\begin{itemize}
  \item Consider the radial coordinate $r$ in such a manner that
  axis is located at $r=0$, then there exists an axially symmetric
  axis. This can be written as
\begin{equation*}
  \mathcal{X}\equiv |\xi^\alpha_{_{(\phi)}}\xi^\beta_{_{(\phi)}}g_{\alpha\beta}|=\mid
  g_{\phi\phi}\mid\rightarrow0,
\end{equation*}
  when $r\rightarrow 0^+$, where $\xi_{_{(\phi)}}$ represents
  the Killing vector.
  \item The spacetime must hold the following condition in order to
  preserve the flatness near the symmetry axis. When $r\rightarrow 0^+$, the condition is
  given by
\begin{equation}\nonumber
  \frac{1}{4\mathcal{X}}\frac{\partial \mathcal{X}}{\partial x^\alpha}\frac{\partial \mathcal{X}}{\partial x^\beta}g^{\alpha\beta}\rightarrow
  1.
  \end{equation}
  \item There must not be any closed timelike curves (rather this is easy to introduce in cylindrical
  spacetimes). We impose the following condition in order to avoid
  these curves
\begin{equation}\nonumber
\xi^\alpha_{_{(\phi)}}\xi^\beta_{_{(\phi)}}g_{\alpha\beta}<0,
\end{equation}
throughout the spacetime.
\item The cylindrical spacetime cannot be asymptotically flat in the axial direction, but must be along radial direction.
\end{itemize}

The exterior region in terms of retarted time coordinates $\upsilon$
and gravitational mass $\mathcal{M}$ is given by \cite{37},
\begin{equation}\label{66}
{{ds^{2}_{(+)}}}=-\left(-\frac{2\mathcal{M}(\upsilon)}{\mathcal{R}}\right)d{\upsilon}^2-
2d{\upsilon}d\mathcal{R}+\mathcal{R}^2\left(d{\phi^2}+\alpha^2d{z}^2\right),
\end{equation}
where $\alpha$ is a constant having dimension $L$. The action of
$f(T)$ gravity \cite{6} is defined as
\begin{equation}\label{1}
{S}=\frac{1}{2{\kappa}}{\int}{{h}}\left({f{(T)}}
+L_{{m}}\right){{d^4{{x}}}},
\end{equation}
here,
${{h}}=\sqrt{-g}={{\textmd{det}}}({{h^a_{\alpha}}}),~h^a_{\alpha}$
represents tetrad components, $ L_{{m}} $ is Lagrangian density of
matter, $f$ is the function of torsion scalar and ${\kappa}=8{\pi}G$
is the coupling constant with $G$ is the gravitational constant. The
variation of action (\ref{1}) with respect to tetrad leads to the
following field equations
\begin{equation}\label{13}
\frac{1}{2}{\kappa}^2{{h}}_{a}^{\hphantom{a}\alpha}{\phi}^{\gamma}_{\alpha}
={{h}_{a}^{\hphantom{a}\alpha}}
{S_{\alpha}^{\hphantom{\alpha}\mu\gamma}}{\partial}_{\mu}{{T}}f_{TT}
+\left[\frac{1}{2}{\partial}_{\mu}
\left({{hh}_{a}^{\hphantom{a}\alpha}}S_{\alpha}^{\hphantom{\alpha}\mu\gamma}\right)
+{{h}_{a}^{\hphantom{a}\alpha}}
{{T}^{\lambda}_{\hphantom{\lambda}\mu\alpha}}S_{\lambda}^{\hphantom{\lambda}\gamma\mu}\right]f_T+
\frac{1}{4}{{h}_{a}^{\hphantom{a}\gamma}}f,
\end{equation}
where $ f_T ,~ f_ {TT} $ stand for first and second order
derivatives with respect to $T$ and torsion scalar is $T\equiv
S_{\sigma}{^{\mu\nu}}T^{\sigma}{_{\mu\nu}}$. The torsion tensor and
superpotential tensor are defined as
\begin{eqnarray}\nonumber
{T^\lambda}_{\mu\nu}=h^{\lambda}_{a}
(\partial_{\nu}h^{a}_{\mu}-\partial_{\mu}h^{a}_{\nu}), \quad
S_{\sigma}{^{\mu\nu}}=\frac{1}{2}(K^{\mu\nu}{_{\sigma}}+\delta^{\mu}_{\sigma}T^{\xi\nu}{_{\xi}}-\delta^{\nu}_{\sigma}T^{\xi\mu}{_{\xi}}),
\end{eqnarray}
where
\begin{eqnarray}
K^{\rho}{_{\mu\nu}}=
\frac{1}{2}(T_{\mu}{^{\rho}}{_{\nu}}+T_{\nu}{^{\rho}}{_{\mu}}-T^{\rho}{_{\mu\nu}}).
\end{eqnarray}

In terms of Einstein tensor, the $f(T)$ field equations take the
following form
\begin{equation}\label{20}
G_{\mu\gamma}=\frac{k^2}{f_{T}}\left({{{\Phi^{\textmd{(M)}}_{\mu\gamma}}}}
+{{{\Phi^{(\textmd{T})}_{\mu\gamma}}}}\right),
\end{equation}
where $\Phi^{\left(\textmd{T}\right)}_{\mu\gamma}$ represents the
torsion part which is defined by
\begin{equation}\label{21}
{{{\Phi^{(\textmd{T})}_{\mu\gamma}}}}=\frac{1}{\kappa^2}\left[-D_{\mu\gamma}f_{TT}
-\frac{1}{4}g_{\mu\gamma}\left({\phi}-Df_{TT}+Rf_{T}\right)\right],\quad
D_{\mu\gamma}={S_{\gamma\mu}}^{\beta}\nabla_{\beta}T.
\end{equation}
The term ${{\Phi^{\textmd{(M)}}_{\mu\gamma}}}$ represents the matter
of the universe. Here we take the matter under consideration as
\begin{equation}\label{32}
{{\Phi^{\textmd{(M)}}_{\mu\gamma}}}={({\rho}+p)}u_{\mu}u_{\gamma}+{pg_{\mu\gamma}}
+{q_{\mu}u_{\gamma}}+{q_{\gamma}u_{\mu}},
\end{equation}
where $\nonumber {u_{\mu}} $ and $\nonumber {q_{\mu}} $ are the
four-velocity and heat conduction satisfy the following relations
$\nonumber {u^{\mu}}={\mathcal{A}^{-1}}{{\delta}^{\mu}_{0}},~
{q^{\mu}}={q\mathcal{B}^{-1}}{{\delta}^{\mu}_{1}},~{q_{\mu}}{u}^{\mu}=0,~{u^{\mu}}{u_{\mu}}=-1,~{\rho}
$ is the energy density of fluid and $p$ denotes the isotropic
pressure.

The selection of vierbein field is crucial for the framework of
$f(T)$ gravity. For cylindrical and spherical geometries, the good
tetrad are non-diagonal tetrads that keep the modification of theory
with no condition applied on torsion scalar. The diagonal tetrads
take the torsion scalar to be constant or vanish in these geometries
named as bad choice of tetrad in $f(T)$ gravity. Therefore, we take
the non-diagonal tetrad for the interior spacetime. In order to
construct non-diagonal tetrad which are suitable for cylindrical
symmetry, different methods are adopted in literature like the local
Lorentz rotations form rotated tetrad and local Lorentz boosts
introduce boosted tetrad. We choose here another approach in which
general coordinate transformation is used \cite{S33}. Consider
static cylindrically symmetric spacetime (\ref{31}). The general
coordinate transformation law is given as
\begin{equation}\label{2.1.2a}
{h^i}_{\mu}=\mathcal{M}^T{h^i}_{\nu'},
\end{equation}
where
$\mathcal{M}^T=\frac{\partial{\mathcal{X}^{\nu'}}}{\partial{\mathcal{X}^{\mu}}}$
denotes the transformation matrix from Cartesian to cylindrical
polar coordinates while $\{\mathcal{X}^{\nu'}\}$ and
$\{\mathcal{X}^{\mu}\}$ are Cartesian and cylindrical polar
coordinates, respectively, i.e.,
$\mathcal{X}^{1'}=r\cos\phi,~\mathcal{X}^{2'}=r\sin\phi$,
$\mathcal{X}^{3'}=z,~
\mathcal{X}^{1}=r,~\mathcal{X}^{2}=\phi,~\mathcal{X}^{3}=z$ and
$\mathcal{X}^{0'}=\mathcal{X}^0=t$. The matrix of transformation
takes the form
\begin{equation}\label{2.1.3}
{h^a}_\mu=\begin{pmatrix}
1 & 0 & 0 & 0 \\
0 & {\cos{\phi}}& {-r\sin{\phi}} &0 \\
0 & {\sin{\phi}} & r{\cos{\phi}} & 0 \\
0       & 0 & 0 & 1
\end{pmatrix}.
\end{equation}
Now we compare the metric (\ref{31}) with Minkowski metric in
cylindrical polar coordinates given as
\begin{equation}\label{2.1.5}
ds^2 =dt^2-dr^2-r^2d\phi^2-dz^2.
\end{equation}
Notice that $g_{00}$ and $g_{11}$ of the metric (\ref{31}) are
obtained by multiplying the corresponding $g_{00}$ and $g_{11}$ of
Minkowski metric by  $\mathcal{A}^2$ and $\mathcal{B}^2$,
respectively and replacing $r$ by $\mathcal{C}$ for $g_{22}$. Thus
multiplying zeroth and first columns of the above matrix by
$\mathcal{A}$ and $\mathcal{B}$, respectively, while replacing $r$
by $\mathcal{C}$, we obtain the tetrad components of static
cylindrically symmetric spacetime
\begin{equation}\label{2.1.33}
{h^a}_\mu=\begin{pmatrix}
\mathcal{A} & 0 & 0 & 0 \\
0 & {{\mathcal{B}~{\cos{\phi}}}}& {{-\mathcal{C}~{\sin{\phi}}}} &0 \\
0 & {{\mathcal{B}~{\sin{\phi}}}} & {{\mathcal{C}~{\cos{\phi}}}} & 0 \\
0       & 0 & 0 & 1
\end{pmatrix}.
\end{equation}

Using the non-zero components of torsion and super-potential
tensors, we get torsion scalar as
\begin{equation}\label{64}
{{T}}={-2}~\left[~\frac{\mathcal{A'}}{\mathcal{A}{{\mathcal{B}^2}}}
~\left(~\frac{{\mathcal{B}}}{{\mathcal{C}}}-\frac{{\mathcal{C}'}}{{\mathcal{C}}}~\right)
+\frac{{\dot{\mathcal{B}}}{\dot{\mathcal{C}}}}{\mathcal{A}^2\mathcal{B}\mathcal{C}}~\right
],
\end{equation}
where dot and prime represent derivatives with respect to time and
radial co-ordinates, respectively. The $f(T)$ field equations of
cylindrically symmetric collapsing star using Eqs.(\ref{31}) and
(\ref{20}) yield
\begin{eqnarray}\label{33}
&&\left(\frac{\mathcal{A}}{{\mathcal{B}}}\right)^2
\left(\frac{{\mathcal{B}'\mathcal{C}'}}{{\mathcal{B}\mathcal{C}}}-\frac{{\mathcal{C}''}}{{\mathcal{C}}}\right)
+\frac{{\dot{\mathcal{B}}}{\dot{{\mathcal{C}}}}}{{\mathcal{B}\mathcal{C}}}=\frac{\kappa^2\mathcal{A}^2}{{f_T}}
\left[{\rho}+\frac{1}{\kappa^2}\left\{\frac{{{T}}f_T-f}{2}
+\frac{1}{2{{\mathcal{B}^2}}}{\times}\right.\right.\notag\\
&&\left.\left.\left(\frac{{\mathcal{B}}}{{\mathcal{C}}}
-\frac{{\mathcal{C}'}}{{\mathcal{C}}}\right){f_T'}\right\}\right].
\\\label{34}
&&\frac{{\dot{\mathcal{C}'}}}{{\mathcal{C}}}-\frac{{\dot{\mathcal{C}}\mathcal{A}'}}
{{\mathcal{C}\mathcal{A}}}-\frac{{\dot{\mathcal{B}}}{\mathcal{C}'}}{{\mathcal{B}\mathcal{C}}}
=\frac{\kappa^2}{f_T} \left\{-{{q\mathcal{A}\mathcal{B}}}
+\frac{{\dot{\mathcal{C}}}}{{2\kappa^2\mathcal{C}}}f_T'\right\}.
\\\label{35}
&&\frac{{\dot{\mathcal{C}'}}}{{\mathcal{C}}}-\frac{{\dot{\mathcal{C}}A'}}
{{\mathcal{C}\mathcal{A}}}-\frac{{\dot{\mathcal{B}}}{{\mathcal{C}'}}}{{\mathcal{B}\mathcal{C}}}
=\frac{\kappa^2}{f_T} \left\{-{{q\mathcal{A}\mathcal{B}}}
+\frac{1}{2\kappa^2}\left(\frac{{\mathcal{C}'}}{{\mathcal{C}}}-\frac{{\mathcal{B}}}{{\mathcal{C}}}\right)
\frac{{\dot{T}}}{{T'}}f_T'\right\}, \\\label{36}
&&\left(\frac{{\mathcal{B}}}{\mathcal{A}}\right)^2\left(\frac{{\dot{A}}{\dot{{\mathcal{C}}}}}{\mathcal{AC}}
-\frac{{\ddot{\mathcal{C}}}}{{\mathcal{C}}}\right )
+\frac{\mathcal{A'\mathcal{C}'}}{\mathcal{AC}}=\frac{{\kappa^2\mathcal{B}^2}}{f_T}\left[{p}
+\frac{1}{\kappa^2}\left\{\frac{f-{{T}}f_T}{2}\right.\right.\notag\\
&&-\left.\left.\frac{{\dot{\mathcal{C}}}{\dot{T}}}{2{\mathcal{A}^2{{\mathcal{C}T'}}}}f_T'\right\}\right].
\\\label{37}
&&\left(\frac{{\mathcal{C}}^2}{\mathcal{A\mathcal{B}}}\right)\left(\frac{\mathcal{A''}}{{\mathcal{B}}}-\frac{{\ddot{\mathcal{B}}}}{\mathcal{A}}
+\frac{{\dot{\mathcal{A}}}{\dot{\mathcal{B}}}}{\mathcal{A}^2}-
\frac{\mathcal{A'\mathcal{B}'}}{{\mathcal{B}^2}}\right)=\frac{{\kappa^2\mathcal{C}^2}}{f_T}\left[{p}+
\frac{1}{\kappa^2}\left\{\frac{f-{{T}}f_T}{2}\right.\right.\notag\\
&&-\left.\left.\left(\frac{{\dot{\mathcal{B}}}{\dot{T}}}{2{\mathcal{A}^2{{\mathcal{B}{T'}}}}}
-{\frac{\mathcal{A'}}{{2\mathcal{A}\mathcal{B}^2}}}\right){f_T}'\right\}\right].
\\\label{38}
&&\frac{\mathcal{A''}}{\mathcal{A}\mathcal{B}^2}-\frac{{\ddot{\mathcal{B}}}}{\mathcal{A}^2{{\mathcal{B}}}}
+\frac{{\dot{A}}{{\dot{\mathcal{B}}}}}{\mathcal{A}^3{{\mathcal{B}}}}-
\frac{\mathcal{A'}\mathcal{B}'}{\mathcal{A}\mathcal{B}^3}+\frac{{\dot{A}}{{\dot{\mathcal{C}}}}}{\mathcal{A}^3{{\mathcal{C}}}}
-\frac{{\ddot{\mathcal{C}}}}{\mathcal{A}^2{{\mathcal{C}}}}-
\frac{{\mathcal{B}'\mathcal{C}'}}{{\mathcal{B}^3}{{\mathcal{C}}}}+\frac{{\mathcal{C}''}}{{\mathcal{B}^2}{{\mathcal{C}}}}\notag\\
&&+\frac{\mathcal{A'\mathcal{C}'}}{\mathcal{A}{{{\mathcal{B}^2}}}{{\mathcal{C}}}}
-\frac{{\dot{\mathcal{B}}}{{\dot{\mathcal{C}}}}}{\mathcal{A}^2{{\mathcal{B}}}{{\mathcal{C}}}}
=\frac{\kappa^2}{f_T}\left[{p}+\frac{1}{\kappa^2}
\left\{\frac{f-{{T}}f_T}{2} -
\left(\frac{{\dot{T}}}{2{\mathcal{A}^2{T'}}}\left(\frac{{\dot{\mathcal{B}}}}{{\mathcal{B}}}
\right.\right.\right.\right.\notag\\
&&+\left.\left.\left.\left.\frac{{\dot{\mathcal{C}}}}{{\mathcal{C}}}\right)
-\frac{1}{{2\mathcal{B}^2}}\left({\frac{\mathcal{A'}}{\mathcal{A}}}
-\frac{{\mathcal{B}}}{{\mathcal{C}}}+\frac{{\mathcal{C}'}}{{\mathcal{C}}}\right)\right){f_T}'\right\}\right].
\end{eqnarray}
Using Eqs.(\ref{34})-(\ref{35}), we get the following relation
\begin{equation}\label{39}
\frac{{\dot{\mathcal{C}}}}{{\mathcal{C}}}=\frac{{\dot{T}}}{{T'}}
\left(\frac{{\mathcal{C}'}}{{\mathcal{C}}}-\frac{{\mathcal{B}}}{{\mathcal{C}}}\right).
\end{equation}

In order to match interior and exterior regions of cylindrical
symmetric collapsing star, we use junction conditions defined by
Darmois. For this purpose, we consider the C-energy, i.e., mass
function representing matter inside the cylinder is given by
\cite{th}
\begin{equation}
m(t,r)=
\frac{\mathcal{L}}{8}\bigg(1-\frac{1}{\mathcal{L}^2}\nabla^\alpha
\hat{r}\nabla_\alpha\hat{r}\bigg)=\mathcal{E}(t,r),
\end{equation}
where $\mathcal{E}$ is the gravitational energy per unit specific
length $\mathcal{L}$ of the cylinder. The areal radius $\hat{r}$ is
defined as $\hat{r}=\mu \mathcal{L}$ where circumference radius has
the relation $\mu^2=\chi_{_{(1)i}}\chi_{_{(1)}}^{i}$ while
$\mathcal{L}^2=\chi_{_{(2)i}}\chi_{_{(2)}}^{i}$. The terms
$\chi_{_{(1)}}=\frac{\partial}{\partial \phi}$ and
$\chi_{_{(2)}}=\frac{\partial}{\partial z}$ are the Killing vectors
corresponding to cylindrical systems. For interior region, the
C-energy turns out
\begin{equation}\label{10}
m(t,r)=\frac{l}{8}\bigg[1+\bigg(\frac{\dot{\mathcal{C}}}{\mathcal{A}}\bigg)^2-\bigg(\frac{\mathcal{C}'}{\mathcal{B}}\bigg)^2\bigg].
\end{equation}
The continuity of Darmois conditions (junction conditions)
establishes the following constraints
\begin{eqnarray}\label{3111}
\frac{\mathcal{L}}{8}\overset{\Sigma^{(e)}}{=}m(t,r)-M, \quad
\mathcal{L}\overset{\Sigma^{(e)}}{=}4C, \quad
-p\overset{\Sigma^{(e)}}{=}\frac{\Phi^{^T}_{11}}{\mathcal{B}^2}
-\frac{\Phi^{^T}_{01}}{\mathcal{A}\mathcal{B}}
\end{eqnarray}
where $\Sigma^{(e)}$ indicates the outer region for measurements
with $r=r_{\Sigma^{(e)}}$=constant. The conservation of total energy
of a system is obtained from Bianchi identities through dynamical
equations.

This relation might be useful in order to simplify the collapse
equation. To analyze the conservation of total energy of a
collapsing star, contracted Bianchi identities in the framework of
$f(T)$ gravity are used. These are defined as
\begin{eqnarray} \label{40}
\left({\Phi^{\textmd{(T)}{\mu\gamma}}}+{\Phi^{\textmd{(M)}\mu\gamma}}\right)_{;\gamma}u_\mu
=\left({\Phi}^{\textmd{(T)})0\gamma}
+{\Phi}^{\textmd{(M)}0\gamma}\right)_{;\gamma}u_0&=&0,
\\\label{45}
\left({\Phi}^{\textmd{(T)}\mu\gamma}+{\Phi}^{\textmd{(M)}\mu\gamma}\right)_{;\gamma}v_\mu
=\left({\Phi}^{\textmd{(T)}1\gamma}
+{\Phi}^{\textmd{(M)}1\gamma}\right)_{;\gamma}v_1&=&0.
\end{eqnarray}
Consequently, we obtain the following two dynamical equations
\begin{eqnarray}
&&{\mathcal{A}^2}\left[\frac{1}{\mathcal{A}^2}\left\{\frac{{{T}}f_T-f}{2}
+\frac{1}{2{{\mathcal{B}^2}}}\left(\frac{{\mathcal{B}}}{{\mathcal{C}}}
-\frac{{\mathcal{C}'}}{{\mathcal{C}}}\right){f_T'}\right\}\right]_{,0}+
{\mathcal{A}^2}\left\{\frac{{\dot{\mathcal{C}}}}{2~{\mathcal{A}^2}
{{\mathcal{B}}^2}{{\mathcal{C}}}}f_T'\right\}_{,1}\notag\nonumber
\\\nonumber&&+\frac{{\dot{\mathcal{A}}}}{\mathcal{A}}\left({{T}}f_T-f\right)
+\frac{1}{{\mathcal{B}\mathcal{C}}}\left(\frac{{\dot{\mathcal{A}}}}{\mathcal{A}}
-\frac{{\dot{A}}{\mathcal{C}'}}{\mathcal{A\mathcal{B}}}
-\frac{3{\mathcal{A'}{\dot{{\mathcal{C}}}}}}{2{\mathcal{A\mathcal{B}}}}+\frac{{\dot{\mathcal{B}}}}
{2{{\mathcal{B}}}}+\frac{{\dot{\mathcal{C}}}}{2{{\mathcal{C}}}}
-\frac{{\dot{\mathcal{B}}}{\mathcal{C}'}}{2{{\mathcal{B}^2}}}+\frac{{\mathcal{B}'}\dot{{\mathcal{C}}}}
{2{{\mathcal{B}^2}}}\right)
\\\nonumber&&+\left(\frac{\mathcal{A'\mathcal{C}'}}{2\mathcal{A}\mathcal{B}^2\mathcal{C}}-\frac{\mathcal{A'}\mathcal{B}}
{2\mathcal{A}\mathcal{B}^2\mathcal{C}}\right)\frac{{\dot{T}}}{{T'}}f_{{T}}'
+{\dot{\rho}}+{q'}{{\mathcal{A}}}+\left(\frac{2{\mathcal{A'}}}{\mathcal{A}}
+\frac{{\mathcal{B}'}}{{\mathcal{B}}}+\frac{{\mathcal{C}'}}{{\mathcal{C}}}\right){q{\mathcal{A}}}
\\&&+\left({{\rho}}+{p}\right)
\left(\frac{{\dot{\mathcal{B}}}}{{\mathcal{B}}}+\frac{{\dot{\mathcal{C}}}}{{\mathcal{C}}}\right)=0.\label{44}
\\\nonumber
&&{{\mathcal{B}^2}}\left\{\frac{{\mathcal{B}^2}}{2\mathcal{A}^2}\left(\frac{{\mathcal{C}'}}{{\mathcal{C}}}
-\frac{{\mathcal{B}}}{{\mathcal{C}}}\right)\frac{{\dot{T}}}
{{T'}}f'_{T}\right\}_{,0}+{{\mathcal{B}^2}}\left\{\frac{1}{{\mathcal{B}^2}}
\left(\frac{f-{{T}{f_T}}}{2}-\frac{{\dot{\mathcal{C}}}{\dot{T}}}
{2{\mathcal{A}^2}{{\mathcal{C}}{T'}}}f'_{T}\right)\right\}_{,1}\notag\\
&&+\frac{{\mathcal{B}'}}{{\mathcal{B}}}\left( f-{{T}}{f'_T}\right)
+\left\{
\left(\frac{3{{\dot{\mathcal{B}}}{\mathcal{C}'}}}{2{\mathcal{A}^2}}{{\mathcal{B}\mathcal{C}}}-\frac{{\dot{\mathcal{B}}}}{\mathcal{A}^2\mathcal{C}}
-\frac{{\mathcal{B}'}{\dot{{\mathcal{C}}}}}{\mathcal{A}^2}{{\mathcal{B}\mathcal{C}}}
+\frac{\mathcal{A'}}{2{\mathcal{A}}{{\mathcal{B}}}{{\mathcal{C}}}}\right.\right.
-\frac{\mathcal{A'}{{\mathcal{C}'}}}{2{\mathcal{A}{{\mathcal{B}^2}{{\mathcal{C}}}}}}\notag\\
&&+\left.\left.\frac{{\dot{\mathcal{A}}}{{\mathcal{C}'}}}{2{\mathcal{A}^3}{{\mathcal{C}}}}
-\frac{{\dot{\mathcal{A}}}{{\mathcal{B}}}}{2{\mathcal{A}^3}{{\mathcal{C}}}}
-\frac{{\mathcal{B}}{{\dot{\mathcal{C}}}}}{2{\mathcal{A}^2}{{\mathcal{C}^2}}}
-\frac{\mathcal{A'}\mathcal{C}'}{2{\mathcal{A}^2}{{\mathcal{C}}}}\right)\frac{{\dot{T}}}{{T'}}
-\frac{\mathcal{A'}\mathcal{C}'}{\mathcal{A}\mathcal{B}^2}{{\mathcal{C}}}
+\frac{{\dot{\mathcal{B}}}{{\dot{\mathcal{C}}}}}{2{\mathcal{A}^2}{{\mathcal{B}\mathcal{C}}}}\right\}{f'_{T}}\notag\\
&&+{p'}+\frac{\dot{q}{\mathcal{A}}}{{\mathcal{B}^2}}+
\left(\frac{3{{\dot{\mathcal{B}}}}}{{\mathcal{B}}}+\frac{{\dot{\mathcal{C}}}}
{{\mathcal{C}}}\right)\frac{q}{\mathcal{A}}{{\mathcal{B}}^2}
+\left({\rho}+{p}\right)\frac{\mathcal{A'}}{\mathcal{A}}=0.\label{49}
\end{eqnarray}

\section{Oscillating States and Collapse Equation}

The selection of $f(T)$ model is very important to establish the
collapse equation representing the instability dynamics of
cylindrically symmetric object in the framework of $f(T)$ gravity.
Here we assume the $f(T)$ model in power-law form up to quadratic
order which is defined by ${f(T)}={T}+{{\chi{T^2}}}$, where $\
{\chi} $ is an arbitrary constant. This model is widely used in the
literature as it indicates the accelerated expansion of the universe
in the phantom phase. The possibility of realistic wormhole
solutions under this model are found. Also, the instability
conditions for a collapsing star under spherically symmetric
collapsing object are discussed. In order to construct the dynamical
equations and explore instability ranges, we assume the linear
perturbation strategy to find the instable behavior. For this we
choose the perturbation scheme in such a way that in the static
configuration (non-perturbed part), metric and matter parts are only
radial dependent while perturbed part contains both radial and time
dependency upto first order where $0<\lambda\leq1$
\cite{26}-\cite{28}. The metric functions under perturbation
strategy become
\begin{eqnarray*}
\mathcal{A}(t,r)=\mathcal{A}_0({r})+\lambda{\hat{a}}(r){\Delta(t)},\quad
{\mathcal{B}(t,r)}={\mathcal{B}_0}({r})+\lambda{\hat{b}}(r){\Delta(t)},\\
{\mathcal{C}(t,r)}={\mathcal{C}_0}({r})+{\lambda}{\hat{c}}(r){\Delta(t)}.
\end{eqnarray*}
The static and perturbed parts of matter components are
\begin{eqnarray*}
{\rho{{(t,r)}}}={\rho_0}{{(r)}}+{\lambda{\overline{\rho}{(t,r)}}},
\quad {p}({t,r})={p}_0({r})+{\lambda{\overline{p}}({t,r})}, \quad
{q}({t,r})={\lambda{\overline{q}}({t,r})},
\end{eqnarray*}
while mass function, torsion scalar and $f(T)$ model are perturbed
as follows
\begin{eqnarray*}
{{m}({t,r})}={{m}_0({r})}+{\lambda{\overline{{m}}}({t,r})},\quad
T(t,r)=T_0(r)+\lambda \omega(r)\Delta(t),\\
{{f(T)}}={{T_0}(1+{\chi}{{T_0}})}+{\omega\lambda\Delta(1+2{\chi}{{T_0}})},\quad
{{f_T(T)}}=1+2{\chi}{{T_0}}+2{\chi}{\lambda}{\Delta}{\omega}.
\end{eqnarray*}
The quantities having zero subscript show the zero order
perturbation. Applying the above set of equations, the perturbed
part of the first dynamical equation is
\begin{eqnarray}\nonumber
&&\left\{{\omega}{\chi}{{T_0}}+\frac{{\chi}{\omega'}\left({{\mathcal{B}_0}-1}\right)}{{\mathcal{C}_0}{{\mathcal{B}^2_o}}}+
\frac{{\chi}{{T'_o}}}{{\mathcal{C}_0}{{\mathcal{B}^2_o}}}\left({{\hat{b}}}-{{\hat{c'}}}+
\frac{{\hat{c}}\left(1-{{\mathcal{B}_0}}\right)}{{\mathcal{C}_0}}-\frac{2{{\hat{b}
\left({{\mathcal{B}_0}-1}\right)}}}{{\mathcal{B}_0}}\right)\right.\notag\\\nonumber
&&\left.-\frac{\hat{a}}{\mathcal{A}_0}\left({\chi}{{T^2_o}}+\frac{2{\chi}{{T'_o}}
\left({{\mathcal{B}_0}-1}\right)}{{\mathcal{C}_0}{{\mathcal{B}_0}^2}}\right)
+{\mathcal{A}^2_0}\left(\frac{\hat{c}{\chi}{{T'_o}}}{{\mathcal{C}_0}{\mathcal{A}^2_o}{{\mathcal{B}^2}_0}}
\right)+\frac{\chi{{T^2_o}}{\hat{a}}}{\mathcal{A}_0}\right.
\\\nonumber
&&\left.+\frac{2{\chi}{{T'_0}}}{{\mathcal{C}_0}{{\mathcal{B}_0}}}
\left(\frac{{\hat{a}}}{\mathcal{A}_0}-
\frac{{\hat{a}}}{\mathcal{A}_0{{\mathcal{B}_0}}}+\frac{{\hat{c}}{\mathcal{A'}_0}}{2{\mathcal{A}_0}{{\mathcal{B}_0}}}
+\frac{{\hat{b}}}{2{{\mathcal{B}_0}}}+\frac{{\hat{c}}}
{2{{\mathcal{C}_0}}}-\frac{{\hat{b}}}{2{{\mathcal{B}^2_o}}}
+\frac{{\hat{c}}{\mathcal{A'}_0}}{{\mathcal{A}_0}{{\mathcal{B}_0}}}\right.\right.\notag\\
&&\left.\left.+\frac{{\hat{c}}{{\mathcal{B}'_o}}}{2{{\mathcal{B}^2_o}}}
+\frac{{\omega}{\mathcal{A'}}\left(1-{{\mathcal{B}_0}}\right){{T'_o}}}
{2{\mathcal{A}_0{{\mathcal{B}_0}}}}\right)\right\}{\dot{\Delta}}
+{\dot{\bar{\rho}}}+{\mathcal{A}_0}{\bar{q'}}+{\bar{q}}{\mathcal{A}_0}
\left(\frac{2{\mathcal{A'}_0}}{\mathcal{A}_0}
+\frac{{\mathcal{B}'}_o}{{\mathcal{B}_0}}\right.\notag\\
&&\left.+\frac{{\mathcal{C}'}_o}{{\mathcal{C}_0}}\right)
+\left({\rho_o}+p_o\right)\left(\frac{2{\hat{b}}}{{\mathcal{B}_0}}
+{\hat{c}}\right){\dot{\Delta}}=0.\label{50}
\end{eqnarray}
Similarly, the second dynamical equation yields
\begin{eqnarray}\nonumber
&&\frac{{\omega}{\chi}{{T^2_o}'}\left(1-{{\mathcal{B}_0}}\right)
{\ddot{\Delta}}}{{\mathcal{C}_0}{\mathcal{A}^2_o}}
+2{{\mathcal{B}_0}{{\hat{b}}}}{\left(\frac{\chi{{T^2_o}}}
{2{{\mathcal{B}^2_o}}}\right)_{,1}}{\Delta}
+{{\mathcal{B}^2_o}}\left\{\frac{1}{{\mathcal{B}^2_o}}\left(\frac{{\hat{b}}
{\chi}{{T^2_o}}}{{\mathcal{B}_0}}
-{\chi}{\omega}{{T_o}}\right)\right\}_{,1}\Delta\notag\\
&&-\frac{2{\chi}{\omega}{{T_o}{{\mathcal{B}'_o}}{\Delta}}}{{\mathcal{B}_0}}
-\frac{1}{{\mathcal{B}_0}}\left({\hat{b}'}
-\frac{{\mathcal{B}'_o}{{\hat{b}}}}{{\mathcal{B}_0}}\right)
{\chi}{{T^2_o}}{\Delta}
+\frac{{\omega}{\chi}{{T^2_o}'}}{{\mathcal{C}_0}{\mathcal{A}_0}}
\left(\frac{\mathcal{A'}_0}{{\mathcal{B}_0}}
-\frac{{\hat{a}}'}{{\mathcal{B}^2_o}}-\frac{{\hat{a}'}}
{\mathcal{A}_0}\right)\notag\\
&&-\frac{2{\omega}'{\chi}{{\hat{a}}'}\Delta}{{\mathcal{C}_0}{\mathcal{A}_0
{{\mathcal{B}^2_o}}}}
-2{\chi}{{T'_o}}\left(\frac{{\hat{a}}}{\mathcal{A}_0}+
\frac{{\hat{b}}}{{\mathcal{B}_0}}
+\frac{{\hat{c}}}{{\mathcal{C}_0}}\right)\Delta
+\dot{\bar{q}}\frac{{\mathcal{B}^2_o}}{\mathcal{A}_0}
+\bar{p}'+\left(\bar{\rho}+\bar{p}\right)\frac{\mathcal{A'}_0}{\mathcal{A}_0}\notag\\
&&+\left({{\rho}_o}+p_o\right)\left(\frac{\hat{a}}{\mathcal{A}_0}\right)'\Delta=0.\nonumber
\end{eqnarray}

The second dynamical equation can be rewritten as
\begin{eqnarray}
&&\textmd{J}_{\textmd{2eq}}+\dot{\bar{q}}\frac{{\mathcal{B}^2_o}}{\mathcal{A}_0}
+\bar{p}'+\left(\bar{\rho}+\bar{p}\right)\frac{\mathcal{A'}_0}{\mathcal{A}_0}
+\left({{\rho}_o}+p_o\right)\left(\frac{\hat{a}}{\mathcal{A}_0}\right)'\Delta=0,\label{51}
\end{eqnarray}
where
\begin{eqnarray}\nonumber
\textmd{J}_{\textmd{2eq}}&=&\frac{{\omega}{\chi}{{T^2_o}'}\left(1-{{\mathcal{B}_0}}\right)
{\ddot{\Delta}}}{{\mathcal{C}_0}{\mathcal{A}^2_0}}
+2{{\mathcal{B}_0}{{\hat{b}}}}{\left(\frac{\chi{{T^2_o}}}
{2{{\mathcal{B}^2_o}}}\right)_{,1}}{\Delta}
+\Delta{{\mathcal{B}^2_o}}\left\{\frac{1}{{\mathcal{B}^2_o}}\left(\frac{{\hat{b}}
{\chi}{{T^2_o}}}{{\mathcal{B}_0}}
-{\chi}{\omega}{{T_o}}\right)\right\}_{,1}\notag\\
&-&\frac{2{\chi}{\omega}{{T_o}{{\mathcal{B}'_o}}{\Delta}}}{{\mathcal{B}_0}}
-\frac{1}{{\mathcal{B}_0}}\left({\hat{b}'}
-\frac{{\mathcal{B}'_o}{{\hat{b}}}}{{\mathcal{B}_0}}\right){\chi}{{T^2_o}}{\Delta}
+\frac{{\omega}{\chi}{{T^2_o}'}}{{\mathcal{C}_0}{\mathcal{A}_0}}\left(\frac{\mathcal{A'}_0}{{\mathcal{B}_0}}
-\frac{{\hat{a}}'}{{\mathcal{B}^2_o}}-\frac{{\hat{a}'}}{\mathcal{A}_0}\right)\notag\\
&-&\frac{2{\omega}'{\chi}{{\hat{a}}'}\Delta}{{\mathcal{C}_0}{\mathcal{A}_0{{\mathcal{B}^2_o}}}}
-2{\chi}{{T'_o}}\left(\frac{{\hat{a}}}{\mathcal{A}_0}+\frac{{\hat{b}}}{{\mathcal{B}_0}}
+\frac{{\hat{c}}}{{\mathcal{C}_0}}\right)\Delta.\nonumber
\end{eqnarray}
The Eq.(\ref{51}) is called as general collapse equation for a
cylindrical symmetric object in $f(T)$ gravity. To discuss the
instability ranges, we have to find the values of
$\overline{\rho},~\overline{p},~\overline{q}$ and $\Delta$. For this
purpose, we do some mathematics in the following. After applying
perturbation on Eq.(\ref{34}), we obtain the following equation
\begin{equation}\nonumber
\left[\left(\frac{\hat{b}}{\mathcal{A}_0{{\mathcal{B}_0}}}\right)'+\left(\frac{\hat{c}}{\mathcal{A}_0}\right)'
+\left(\frac{1}{{\mathcal{C}_0}}+\frac{{\mathcal{B}'_o}}{{\mathcal{B}_0}}\right)
\left(\frac{{\hat{c}}}{\mathcal{A}_0}\right)\right]\dot{\Delta}
=\left[\frac{{\hat{c}}}{{2\mathcal{C}_0}}\left(1+2\chi{{T_o}}\right)\right]\dot{\Delta}
+8\pi{{\mathcal{B}^2_o}}\bar{q},
\end{equation}
which yields
\begin{equation}
\bar{q}=\frac{\left[\left(\frac{\hat{b}}{\mathcal{A}_0{{\mathcal{B}_0}}}\right)'
+\left(\frac{\hat{c}}{\mathcal{A}_0}\right)'
+\left(\frac{1}{{\mathcal{C}_0}}+\frac{{\mathcal{B}'_o}}{{\mathcal{B}_0}}\right)
\left(\frac{{\hat{c}}}{\mathcal{A}_0}\right)
-\frac{{\hat{c}}}{{2\mathcal{C}_0}}\left(1+2\chi{{T_o}}\right)\right]
\dot{\Delta}}{8\pi{{\mathcal{B}^2_o}}}.\label{52}
\end{equation}

Putting the above value of ${\bar{q}}$ in Eq.(\ref{50}), we get
\begin{eqnarray}\nonumber
&&\left\{{\omega}{\chi}{{T_0}}+\frac{{\chi}{\omega'}\left({{\mathcal{B}_0}-1}\right)}
{{\mathcal{C}_0}{{\mathcal{B}^2_o}}}+
\frac{{\chi}{{T'_o}}}{{\mathcal{C}_0}{{\mathcal{B}^2_o}}}
\left({{\hat{b}}}-{{\hat{c'}}}+
\frac{{\hat{c}}\left(1-{{\mathcal{B}_0}}\right)}{{\mathcal{C}_0}}-\frac{2{{\hat{b}
\left({{\mathcal{B}_0}-1}\right)}}}{{\mathcal{B}_0}}\right)\right.\notag\\
&&\left.-\frac{\hat{a}}{\mathcal{A}_0}\left({\chi}{{T^2_o}}
+\frac{2{\chi}{{T'_o}}
\left({{\mathcal{B}_0}-1}\right)}{{\mathcal{C}_0}{{\mathcal{B}_0}^2}}\right)
+{\mathcal{A}^2_0}\left(\frac{\hat{c}{\chi}{{T'_o}}}
{{\mathcal{C}_0}{\mathcal{A}^2_o}{{\mathcal{B}^2}_o}}
\right)+\frac{\chi{{T^2_o}}{\hat{a}}}{\mathcal{A}_0}\right.\notag\\
&&\left.+\frac{2{\chi}{{T'_0}}}{{\mathcal{C}_0}{{\mathcal{B}_0}}}
\left(\frac{{\hat{a}}}{\mathcal{A}_0}-
\frac{{\hat{a}}}{\mathcal{A}_0{{\mathcal{B}_0}}}+\frac{{\hat{c}}
{\mathcal{A'}_0}}{2{\mathcal{A}_0}{{\mathcal{B}_0}}}
+\frac{{\hat{b}}}{2{{\mathcal{B}_0}}}+\frac{{\hat{c}}}{2{{\mathcal{C}_0}}}
-\frac{{\hat{b}}}{2{{\mathcal{B}^2_o}}}
+\frac{{\hat{c}}{\mathcal{A'}_0}}{{\mathcal{A}_0}{{\mathcal{B}_0}}}\right.\right.\notag\\
&&\left.\left.+\frac{{\hat{c}}{{\mathcal{B}'_o}}}{2{{\mathcal{B}^2_o}}}
+\frac{{\omega}{\mathcal{A'}}\left(1-{{\mathcal{B}_0}}\right){{T'_o}}}{2{\mathcal{A}_0
{{\mathcal{B}_0}}}}\right)\right\}{\dot{\Delta}}
+{\dot{\bar{\rho}}}+\left({\rho_o}+p_o\right)\left(\frac{2{\hat{b}}}{{\mathcal{B}_0}}+{\hat{c}}\right)
{\dot{\Delta}}\notag\\
&&+\left[\left(\frac{\hat{b}}{\mathcal{A}_0{{\mathcal{B}_0}}}\right)'
+\left(\frac{\hat{c}}{\mathcal{A}_0}\right)'
+\left(\frac{1}{{\mathcal{C}_0}}+\frac{{\mathcal{B}'_o}}{{\mathcal{B}_0}}\right)
\left(\frac{{\hat{c}}}{\mathcal{A}_0}\right)
-\frac{{\hat{c}}}{{2\mathcal{C}_0}}\left(1+2\chi{{T_o}}\right)
\right]'\notag\\
&&\times\frac{\mathcal{A}_0\dot{\Delta}}{8\pi{{{\mathcal{B}'_o}^2}}}
+{\frac{\left[\left(\frac{\hat{b}}{\mathcal{A}_0{{\mathcal{B}_0}}}\right)'
+\left(\frac{\hat{c}}{\mathcal{A}_0}\right)'
+\left(\frac{1}{{\mathcal{C}_0}}+\frac{{\mathcal{B}'_o}}{{\mathcal{B}_0}}\right)
\left(\frac{{\hat{c}}}{\mathcal{A}_0}\right)
-\frac{{\hat{c}}}{{2\mathcal{C}_0}}\left(1+2\chi{{T_o}}\right)\right]
}{8\pi{{\mathcal{B}^2_o}}}}\notag\\
&&{\times}{\mathcal{A}_0}\dot{\Delta}
\left(\frac{2{\mathcal{A'}_0}}{\mathcal{A}_0}
+\frac{{\mathcal{B}'}_o}{{\mathcal{B}_0}}
+\frac{{\mathcal{C}'}_o}{{\mathcal{C}_0}}\right) =0.\label{53}
\end{eqnarray}
We can write this equation as
\begin{eqnarray}\nonumber
&&{\mathcal{A}_0}\left({\frac{\left(\frac{\hat{b}}{\mathcal{A}_0{{\mathcal{B}_0}}}\right)'
+\left(\frac{\hat{c}}{\mathcal{A}_0}\right)'
+\left(\frac{1}{{\mathcal{C}_0}}+\frac{{\mathcal{B}'_o}}{{\mathcal{B}_0}}\right)
\left(\frac{{\hat{c}}}{\mathcal{A}_0}\right)
-\frac{{\hat{c}}}{{2\mathcal{C}_0}}\left(1+2\chi{{T_o}}\right)}
{8\pi{{\mathcal{B}^2_o}}}}\right)'\dot{\Delta}\notag\\
&&+{\frac{\left(\left(\frac{\hat{b}}{\mathcal{A}_0{{\mathcal{B}_0}}}\right)'
+\left(\frac{\hat{c}}{\mathcal{A}_0}\right)'
+\left(\frac{1}{{\mathcal{C}_0}}+\frac{{\mathcal{B}'_o}}{{\mathcal{B}_0}}\right)
\left(\frac{{\hat{c}}}{\mathcal{A}_0}\right)
-\frac{{\hat{c}}}{{2\mathcal{C}_0}}\left(1+2\chi{{T_o}}\right)\right)
\dot{\Delta}}{8\pi{{\mathcal{B}^2_o}}}}{\times}\notag\\
&&{\mathcal{A}_0} \left(\frac{2{\mathcal{A'}_0}}{\mathcal{A}_0}
+\frac{{\mathcal{B}'}_o}{{\mathcal{B}_0}}
+\frac{{\mathcal{C}'}_o}{{\mathcal{C}_0}}\right)
+\left({\rho_o}+p_o\right)\left(\frac{2{\hat{b}}}{{\mathcal{B}_0}}+{\hat{c}}\right){\dot{\Delta}}
+\textmd{J}_{\textmd{1eq}}{\dot{\Delta}}+{\dot{\bar{\rho}}}=0,\notag\\\label{54}
\end{eqnarray}
where
\begin{eqnarray}\nonumber
\textmd{J}_{\textmd{1eq}}&=&{\omega}{\chi}{{T_0}}+\frac{{\chi}{\omega'}({{\mathcal{B}_0}-1})}
{{\mathcal{C}_0}{{\mathcal{B}^2_o}}}+
\frac{{\chi}{{T'_o}}}{{\mathcal{C}_0}{{\mathcal{B}^2_o}}}\bigg({{\hat{b}}}
-{{\hat{c'}}}+
\frac{{\hat{c}}(1-{{\mathcal{B}_0}})}{{\mathcal{C}_0}}-\frac{2{{\hat{b}
({{\mathcal{B}_0}-1})}}}{{\mathcal{B}_0}}\bigg)\\\nonumber
&-&\frac{\hat{a}}{\mathcal{A}_0}\bigg({\chi}{{T^2_o}}+\frac{2{\chi}{{T'_o}}
({{\mathcal{B}_0}-1})}{{\mathcal{C}_0}{{\mathcal{B}_0}^2}}\bigg)
+{\mathcal{A}^2_0}\bigg(\frac{\hat{c}{\chi}{{T'_o}}}{{\mathcal{C}_0}{\mathcal{A}^2_o}{{\mathcal{B}^2}_o}}\bigg)
+\frac{\chi{{T^2_o}}{\hat{a}}}{\mathcal{A}_0}\\\nonumber
&+&\frac{2{\chi}{{T'_0}}}{{\mathcal{C}_0}{{\mathcal{B}_0}}}
\bigg(\frac{{\hat{a}}}{\mathcal{A}_0}-
\frac{{\hat{a}}}{\mathcal{A}_0{{\mathcal{B}_0}}}+
\frac{{\hat{c}}{\mathcal{A'}_0}}{2{\mathcal{A}_0}{{\mathcal{B}_0}}}
+\frac{{\hat{b}}}{2{{\mathcal{B}_0}}}+\frac{{\hat{c}}}{2{{\mathcal{C}_0}}}
-\frac{{\hat{b}}}{2{{\mathcal{B}^2_o}}}
+\frac{{\hat{c}}{\mathcal{A'}_0}}{{\mathcal{A}_0}{{\mathcal{B}_0}}}\\\label{ce}
&+&\frac{{\hat{c}}{{\mathcal{B}'_o}}}{2{{\mathcal{B}^2_o}}}
+\frac{{\omega}{\mathcal{A'}}(1-{{\mathcal{B}_0}}){{T'_o}}}
{2{\mathcal{A}_0{{\mathcal{B}_0}}}}\bigg).
\end{eqnarray}
Integrating the above equation w.r.t "$t$", we get
\begin{eqnarray}\nonumber
&&
{\mathcal{A}_0}\left({\frac{\left(\frac{\hat{b}}{\mathcal{A}_0{{\mathcal{B}_0}}}\right)'
+\left(\frac{\hat{c}}{\mathcal{A}_0}\right)'
+\left(\frac{1}{{\mathcal{C}_0}}+\frac{{\mathcal{B}'_o}}{{\mathcal{B}_0}}\right)
\left(\frac{{\hat{c}}}{\mathcal{A}_0}\right)
-\frac{{\hat{c}}}{{2\mathcal{C}_0}}\left(1+2\chi{{T_o}}\right)}
{8\pi{{\mathcal{B}^2_o}}}}\right)'{\Delta}\notag\\
&&+{\frac{\left(\left(\frac{\hat{b}}{\mathcal{A}_0{{\mathcal{B}_0}}}\right)'
+\left(\frac{\hat{c}}{\mathcal{A}_0}\right)'
+\left(\frac{1}{{\mathcal{C}_0}}+\frac{{\mathcal{B}'_o}}{{\mathcal{B}_0}}\right)
\left(\frac{{\hat{c}}}{\mathcal{A}_0}\right)
-\frac{{\hat{c}}}{{2\mathcal{C}_0}}\left(1+2\chi{{T_o}}\right)\right)
{\Delta}}{8\pi{{\mathcal{B}^2_o}}}}{\times}\notag\\
&&{\mathcal{A}_0} \left(\frac{2{\mathcal{A'}_0}}{\mathcal{A}_0}
+\frac{{\mathcal{B}'}_o}{{\mathcal{B}_0}}
+\frac{{\mathcal{C}'}_o}{{\mathcal{C}_0}}\right)
+\left({\rho_o}+p_o\right)\left(\frac{2{\hat{b}}}{{\mathcal{B}_0}}+{\hat{c}}\right){\Delta}
+\textmd{J}_{\textmd{1eq}}\Delta+{\bar{\rho}}=0.\label{55}
\end{eqnarray}

The Harrison-Wheeler equation of state defined by
\begin{equation}\label{30}
{\overline{p}}={\Gamma}\frac{p_0}{\rho_0+p_0}{\overline{\rho}},
\end{equation}
where ${\Gamma}$ is named as adiabatic index. The adiabatic index
defines the instability ranges of a self-gravitating collapsed
objects. Here we discuss these ranges for cylindrically symmetric
self gravitational fluid in the frame-work of $f(T)$ gravity both in
Newtonian and post Newtonian Regimes through adiabatic index under
collapse equation. Equation (\ref{55}) can be written as
\begin{eqnarray}\nonumber
{\bar{\rho}}&=&-\bigg[
{\mathcal{A}_0}\bigg(\frac{(\frac{\hat{b}}{\mathcal{A}_0{{\mathcal{B}_0}}})'
+(\frac{\hat{c}}{\mathcal{A}_0})'
+(\frac{1}{{\mathcal{C}_0}}+\frac{{\mathcal{B}'_o}}{{\mathcal{B}_0}})
(\frac{{\hat{c}}}{\mathcal{A}_0})
-\frac{{\hat{c}}}{{2\mathcal{C}_0}}(1+2\chi{{T_o}})} {8\pi
{{\mathcal{B}^2_0}}}\bigg)'\Delta\\\nonumber
&+&\frac{(\frac{\hat{b}}{\mathcal{A}_0{{\mathcal{B}_0}}})'
+(\frac{\hat{c}}{\mathcal{A}_0})'
+(\frac{1}{{\mathcal{C}_0}}+\frac{{\mathcal{B}'_o}}{{\mathcal{B}_0}})
(\frac{{\hat{c}}}{\mathcal{A}_0})
-\frac{\hat{c}}{{2\mathcal{C}_0}}(1+2\chi{{T_o}})}{8\pi
{{\mathcal{B}^2_0}}}\Delta\\\nonumber &\times&{\mathcal{A}_0}
\bigg(\frac{2{\mathcal{A'}_0}}{\mathcal{A}_0}
+\frac{{\mathcal{B}'}_o}{{\mathcal{B}_0}}
+\frac{{\mathcal{C}'}_o}{{\mathcal{C}_0}}\bigg)+\textmd{J}_{\textmd{1eq}}\Delta
+({\rho_o}+p_o)(\frac{2{\hat{b}}}{{\mathcal{B}_0}}+{\hat{c}})
{\Delta}\bigg].\label{56}
\end{eqnarray}
To find the value of ${\overline{p}}$, we insert the value of
$\bar{\rho}$ in Eq.(\ref{30}), which gives
\begin{eqnarray}\nonumber
{\overline{p}}&=&{\Gamma}\frac{-p_0}{\rho_0+p_0} \bigg[
\bigg({\frac{(\frac{\hat{b}}{\mathcal{A}_0{{\mathcal{B}_0}}})'
+(\frac{\hat{c}}{\mathcal{A}_0})'
+(\frac{1}{{\mathcal{C}_0}}+\frac{{\mathcal{B}'_o}}{{\mathcal{B}_0}})
(\frac{{\hat{c}}}{\mathcal{A}_0})
-\frac{{\hat{c}}}{{2\mathcal{C}_0}}(1+2\chi{{T_o}})}
{8\pi{{\mathcal{B}^2_o}}}}\bigg)'\\\nonumber
&\times&{\Delta}{\mathcal{A}_0}+{\frac{(\frac{\hat{b}}{\mathcal{A}_0{{\mathcal{B}_0}}})'
+(\frac{\hat{c}}{\mathcal{A}_0})'
+(\frac{1}{{\mathcal{C}_0}}+\frac{{\mathcal{B}'_o}}{{\mathcal{B}_0}})
(\frac{{\hat{c}}}{\mathcal{A}_0})
-\frac{{\hat{c}}}{{2\mathcal{C}_0}}(1+2\chi{{T_o}})}{8\pi{{\mathcal{B}^2_o}}}}\\\label{57}
&{\times}&{\Delta}{\mathcal{A}_0}\bigg(\frac{2{\mathcal{A'}_0}}{\mathcal{A}_0}
+\frac{{\mathcal{B}'}_o}{{\mathcal{B}_0}}
+\frac{{\mathcal{C}'}_o}{{\mathcal{C}_0}}\bigg)
+({\rho_o}+p_o)(\frac{2{\hat{b}}}{{\mathcal{B}_0}}+{\hat{c}}){\Delta}
+\textmd{J}_{\textmd{1eq}}\Delta\bigg].
\end{eqnarray}
In order to determine the value of $ {\Delta(t)}$, we perturbed
Eq.(\ref{36}) which is given by
\begin{eqnarray}
&&\frac{-{\hat{c}}}{{\mathcal{C}_0}{\mathcal{A}^2_0}}\ddot{\Delta}
+\frac{1}{{\mathcal{C}_0}{\mathcal{A}_0{{\mathcal{B}^2_0}}}}
\left(\frac{-2{{\hat{b}{\mathcal{A'}_0}}}}{{\mathcal{B}_0}}
-\frac{{\hat{a}}}{\mathcal{A}_0}{\mathcal{A'}_0}
-\frac{{\hat{c}}}{{\mathcal{C}_0}}{\mathcal{A'}_0}+{{\hat{a}'}}
+{{\hat{c}'}}{\mathcal{A}_0'}
\right)\Delta\notag\\
&&=-\frac{2{\chi}{\omega}{\Delta}p_0}{(1+2{\chi}{{T_o}})^2}+\frac{\overline{p}}{1+2{\chi}{{T_o}}}-\frac{\omega
\chi
T_o\Delta}{1+2{\chi}{{T_o}}}-\frac{\omega\chi^2T^2_o\Delta}{(1+2{\chi}{{T_o}})^2},\label{58}
\end{eqnarray}
The solution of this equation can be found through matching of
interior and exterior regions which leads $r=r_{\Sigma}=
\textmd{constant}~r$, on boundary surface. Under this condition and
using value of ${\overline{p}}$, Eq.(\ref{58}) yields
\begin{equation}\label{t}
\ddot{\Delta}-{\psi}_{_{\Sigma}}{\Delta}\overset{\Sigma}{=}0,
\end{equation}
where
\begin{equation}\nonumber
{\psi}_{_{\Sigma}}=\frac{2{\mathcal{C}_{_{\Sigma}}}
{\chi}{\omega}{\mathcal{A}^2_{_{\Sigma}}}p_{_{\Sigma}}}{\hat{c}}-\frac{\mathcal{C}_{_{\Sigma}}\mathcal{A}^2_{_{\Sigma}}
\overline{p}_{_{\Sigma}}}{\hat{c}(1+2{\chi}{{T_{_{\Sigma}}}})}+\frac{\omega\chi\mathcal{C}_{_{\Sigma}}\mathcal{A}^2_{_{\Sigma}}
T_{_{\Sigma}}}{\hat{c}(1+2{\chi}{{T_{_{\Sigma}}}})}+\frac{\omega\chi^2\mathcal{C}_{_{\Sigma}}\mathcal{A}^2_{_{\Sigma}}
T^2_{_{\Sigma}}}{\hat{c}(1+2{\chi}{{T_{_{\Sigma}}}})^2}.
\end{equation}
In the above equation, all the terms with subscript $\Sigma$ are as
follows.
\begin{eqnarray*}
T_{_{\Sigma}}=T_{o}(r_{_{\Sigma}}),~\mathcal{C}_{_{\Sigma}}=\mathcal{C}_0(r_{_{\Sigma}}),~\mathcal{A}_{_{\Sigma}}=\mathcal{A}_0(r_{_{\Sigma}}),
~\mathcal{B}_{_{\Sigma}}=\mathcal{B}_0(r_{_{\Sigma}}),~
p_{_{\Sigma}}=p_0(r_{_{\Sigma}}),\\
\overline{p}_{_{\Sigma}}=\overline{p}(r_{_{\Sigma}})=(\rho_{_{\Sigma}}+p_{_{\Sigma}})(\frac{2\hat{b}}{\mathcal{B}_{_{\Sigma}}}+\hat{c})+
\omega\chi T_{_{\Sigma}},~\rho_{_{\Sigma}}=\rho_0(r_{_{\Sigma}}).
\end{eqnarray*}
The solution of Eq.(\ref{t}) is
\begin{equation}\nonumber
{\Delta}\left({{t}}\right)={a_{1}}{\exp}^{\sqrt{{\psi}_{_{\Sigma}}}}t
+{a_{2}}{\exp}^{-\sqrt{{\psi}_{_{\Sigma}}}}t.
\end{equation}
This solution exhibits the stability as well as instability
configurations and ${a_1}$ and ${a_2}$ are constants. Here to
discuss the instability analysis, we take only static solution of
cylindrical collapsing star which implies $ a_2=0 $, while
${a_1}=-1$. We get
\begin{equation}\label{23+}
\Delta=-{\exp}^{\sqrt{{\psi}_{_{\Sigma}}}}t, \quad\psi_{_{\Sigma}}>0
\end{equation}

Putting all the correspondent values in general collapse equation,
we attain the collapse equation of cylindrically symmetric object in
$f(T)$ gravity as
\begin{eqnarray}
&&{\frac{\left(\frac{\hat{b}}{\mathcal{A}_0{{\mathcal{B}_0}}}\right)'
+\left(\frac{\hat{c}}{\mathcal{A}_0}\right)'
+\left(\frac{1}{{\mathcal{C}_0}}+\frac{{\mathcal{B}'_o}}{{\mathcal{B}_0}}\right)
\left(\frac{{\hat{c}}}{\mathcal{A}_0}\right)
-\frac{{\hat{c}}}{{2\mathcal{C}_0}}\left(1+2\chi{{T_o}}\right)
}{8\pi{{\mathcal{B}^2_o}}}}
\frac{{\mathcal{B}^2_o}}{\mathcal{A}_0}\ddot{\Delta}+J_{2eq}
\notag\\
&+&
\left[-{\mathcal{A}_0}\left({\frac{\left(\frac{\hat{b}}{\mathcal{A}_0{{\mathcal{B}_0}}}\right)'
+\left(\frac{\hat{c}}{\mathcal{A}_0}\right)'
+\left(\frac{1}{{\mathcal{C}_0}}+\frac{{\mathcal{B}'_o}}{{\mathcal{B}_0}}\right)
\left(\frac{{\hat{c}}}{\mathcal{A}_0}\right)
-\frac{{\hat{c}}}{{2\mathcal{C}_0}}\left(1+2\chi{{T_o}}\right)}
{8\pi{{\mathcal{B}^2_o}}}}\right)'\right.\notag\\
&-&\left.\left.{\frac{\left(\frac{\hat{b}}{\mathcal{A}_0{{\mathcal{B}_0}}}\right)'
+\left(\frac{\hat{c}}{\mathcal{A}_0}\right)'
+\left(\frac{1}{{\mathcal{C}_0}}+\frac{{\mathcal{B}'_o}}{{\mathcal{B}_0}}\right)
\left(\frac{{\hat{c}}}{\mathcal{A}_0}\right)
-\frac{{\hat{c}}}{{2\mathcal{C}_0}}\left(1+2\chi{{T_o}}\right)}
{8\pi{{\mathcal{B}^2_o}}}}{\mathcal{A}_0}\right.\right.\notag\\
&{\times}&\left. \left(\frac{2{\mathcal{A'}_0}}{\mathcal{A}_0}
+\frac{{\mathcal{B}'}_o}{{\mathcal{B}_0}}
+\frac{{\mathcal{C}'}_o}{{\mathcal{C}_0}}\right)
-\left({\rho_o}+p_o\right)\left(\frac{2{\hat{b}}}{{\mathcal{B}_0}}+{\hat{c}}
\right)-2J_{1eq}-\frac{{\Gamma}p_0{\mathcal{A}_0}}{\rho_0+p_0} \right.\notag\\
&\times&\left.\left\{
\left({\frac{\left(\frac{\hat{b}}{\mathcal{A}_0{{\mathcal{B}_0}}}\right)'
+\left(\frac{\hat{c}}{\mathcal{A}_0}\right)'
+\left(\frac{1}{{\mathcal{C}_0}}+\frac{{\mathcal{B}'_o}}{{\mathcal{B}_0}}\right)
\left(\frac{{\hat{c}}}{\mathcal{A}_0}\right)
-\frac{{\hat{c}}}{{2\mathcal{C}_0}}\left(1+2\chi{{T_o}}\right)}
{8\pi{{\mathcal{B}^2_o}}}}\right)'\right.\right.\notag\\
&-&\left.\left.{\frac{\left(\frac{\hat{b}}{\mathcal{A}_0{{\mathcal{B}_0}}}\right)'
+\left(\frac{\hat{c}}{\mathcal{A}_0}\right)'
+\left(\frac{1}{{\mathcal{C}_0}}+\frac{{\mathcal{B}'_o}}{{\mathcal{B}_0}}\right)
\left(\frac{{\hat{c}}}{\mathcal{A}_0}\right)
-\frac{{\hat{c}}}{{2\mathcal{C}_0}}\left(1+2\chi{{T_o}}\right)}
{8\pi{{\mathcal{B}^2_o}}}}{\mathcal{A}_0}\right.\right.\notag\\
&{\times}&\left.\left. \left(\frac{2{\mathcal{A'}_0}}{\mathcal{A}_0}
+\frac{{\mathcal{B}'}_o}{{\mathcal{B}_0}}
+\frac{{\mathcal{C}'}_o}{{\mathcal{C}_0}}\right)
-\left({\rho_o}+p_o\right)\left(\frac{2{\hat{b}}}{{\mathcal{B}_0}}+{\hat{c}}\right)
\right\}\right]
\left(\frac{\mathcal{A'}_0}{\mathcal{A}_0}\right){\Delta}+\left[\frac{-{\Gamma}p_0}{\rho_0+p_0}
\notag\right.\\
&\times&\left\{{\mathcal{A}_0}\left.
\left({\frac{\left(\frac{\hat{b}}{\mathcal{A}_0{{\mathcal{B}_0}}}\right)'
+\left(\frac{\hat{c}}{\mathcal{A}_0}\right)'
+\left(\frac{1}{{\mathcal{C}_0}}+\frac{{\mathcal{B}'_o}}{{\mathcal{B}_0}}\right)
\left(\frac{{\hat{c}}}{\mathcal{A}_0}\right)
-\frac{{\hat{c}}}{{2\mathcal{C}_0}}\left(1+2\chi{{T_o}}\right)}
{8\pi{{\mathcal{B}^2_o}}}}\right)'\right.\right.\notag\\
&+&\left.\left.\textmd{J}_{\textmd{1eq}}+{\frac{\left(\frac{\hat{b}}{\mathcal{A}_0{{\mathcal{B}_0}}}\right)'
+\left(\frac{\hat{c}}{\mathcal{A}_0}\right)'
+\left(\frac{1}{{\mathcal{C}_0}}+\frac{{\mathcal{B}'_o}}{{\mathcal{B}_0}}\right)
\left(\frac{{\hat{c}}}{\mathcal{A}_0}\right)
-\frac{{\hat{c}}}{{2\mathcal{C}_0}}\left(1+2\chi{{T_o}}\right)
}{8\pi{{\mathcal{B}^2_o}}}}\right.\right.\notag\\
&{\times}&\left.\left.{\mathcal{A}_0}
\left(\frac{2{\mathcal{A'}_0}}{\mathcal{A}_0}
+\frac{{\mathcal{B}'}_o}{{\mathcal{B}_0}}
+\frac{{\mathcal{C}'}_o}{{\mathcal{C}_0}}\right)
+\left({\rho_o}+p_o\right)\left(\frac{2{\hat{b}}}{{\mathcal{B}_0}}+{\hat{c}}\right)\right\}\right]'\Delta
+\left({{\rho}_o}+p_o\right)\notag\\
&\times&\left(\frac{\hat{a}}{\mathcal{A}_0}\right)'\Delta=0,\label{60}
\end{eqnarray}
where $\Delta$ is given in Eq.(\ref{23+}).

\section{Instability Ranges}

Now we consider the collapse equation (\ref{60}) with Newtonian and
Post Newtonian regime constraints to figure out the instability
ranges of self gravitating object in $f(T)$ gravity.

\subsection{Newtonian Order}

The constraints for Newtonian regime are given as follows
\begin{eqnarray}\nonumber
&&{\mathcal{A}_0}=1={{\mathcal{B}_0}},\quad \frac{p_o}{\rho_o}<1,\notag\\
\Rightarrow&&{\mathcal{A'}_0}=0={{\mathcal{B}'_0}}, \quad
\frac{\frac{p_o}{\rho_o}}
{1+\frac{p_o}{\rho_o}}\rightarrow{0}.\nonumber\
\end{eqnarray}
Using these conditions in Eq.(\ref{60}), the collapse equation turns
out as
\begin{eqnarray}
\Gamma
p'_o(2\hat{b}+\hat{c})'=(\rho_o+p_o)\hat{a}'+\frac{1}{\Delta}{\textmd{J}}^{\textmd{N}}_{\textmd{2eq}}+\bigg(\hat{b}'+\hat{c}'
+\frac{\hat{c}}{\mathcal{C}_0}-\frac{\hat{c}\xi
T_0}{\mathcal{C}_0}\bigg)\psi_{\Sigma},\label{62}
\end{eqnarray}
which indicates the hydrostatic state of a cylindrically symmetric
self gravitating fluid where
${\textmd{J}}^{\textmd{N}}_{\textmd{2eq}}$ represents those terms in
${\textmd{J}}_{\textmd{2eq}}$ which comes through Newtonian
approximation. This equation contains the matter, metric and torsion
scalar components which take part to develop the instability ranges.
It is noted that we take adiabatic index with positive sign all over
the scenario to preserve the variation between gravitational forces,
gradient of heat and pressure components. This system will be
unstable if
\begin{eqnarray}
&&\Gamma<\frac{(\rho_o+p_o)\hat{a}'-{\textmd{e}}^{-\sqrt{\psi_{\Sigma}}t}{\textmd{J}}^{\textmd{N}}_{\textmd{2eq}}+\bigg(\hat{b}'+\hat{c}'
+\frac{\hat{c}}{\mathcal{C}_0}-\frac{\hat{c}\xi
T_0}{\mathcal{C}_0}\bigg)\psi_{\Sigma}}{p'_o(2\hat{b}+\hat{c})'}.\label{63}
\end{eqnarray}
The left hand side of inequality remains positive while the system
remains in instable state till this inequality holds. Now we discuss
the following three cases.
\begin{itemize}
  \item$\textbf{Case 1}$:\end{itemize}
If the term
$(\rho_o+p_o)\hat{a}'+\frac{1}{\Delta}{\textmd{J}}^{\textmd{N}}_{\textmd{2eq}}+\bigg(\hat{b}'+\hat{c}'
+\frac{\hat{c}}{\mathcal{C}_0}-\frac{\hat{c}\xi
T_0}{\mathcal{C}_0}\bigg)\psi_{\Sigma}$ balanced by the term
$p'_o(2\hat{b}+\hat{c})'$ in Eq.(\ref{62}), then we obtain
$\Gamma=1$. This leads to the hydrostatic equilibrium state for this
particular case for cylindrically symmetric self gravitating object.
\begin{itemize}
  \item$\textbf{Case 2}  $:\end{itemize}
Now if
$(\rho_o+p_o)\hat{a}'+\frac{1}{\Delta}{\textmd{J}}^{\textmd{N}}_{\textmd{2eq}}+\bigg(\hat{b}'+\hat{c}'
+\frac{\hat{c}}{\mathcal{C}_0}-\frac{\hat{c}\xi
T_0}{\mathcal{C}_0}\bigg)\psi_{\Sigma}$ is lesser than
$p'_o(2\hat{b}+\hat{c})'$, it leads to $\Gamma<1$ through
Eq.(\ref{62}). The constraint on adiabatic index, $0<\Gamma<1$
contributes to the instable state of the fluid without collapse.
\begin{itemize}
  \item$ \textbf{Case 3}$:\end{itemize}
Eq.(\ref{62}) implies that if $p'_o(2\hat{b}+\hat{c})'$ is lesser
than
$(\rho_o+p_o)\hat{a}'+\frac{1}{\Delta}{\textmd{J}}^{\textmd{N}}_{\textmd{2eq}}+\bigg(\hat{b}'+\hat{c}'
+\frac{\hat{c}}{\mathcal{C}_0}-\frac{\hat{c}\xi
T_0}{\mathcal{C}_0}\bigg)\psi_{\Sigma}$, we get $\Gamma>1$ which
establishes instability ranges for collapse of the self-gravitating
cylindrically symmetric object in the framework of $f(T)$ gravity.

It is noted that, we may recover the general relativity in Newtonian
limit for instability range as $\Gamma<\frac{4}{3}$.

\subsection{Post-Newtonian Order}

In the post-Newtonian order, we deal the dynamics of cylindrical
symmetric star as $ \mathcal{A}_0=1-\frac{m_o}{r},~
{\mathcal{B}_0}=1+\frac{m_o}{r} $. We have taken $
\left\{\frac{m}{r_0}\right\} $ terms upto first order and neglect
all the higher terms. Using these constraints, we obtain some terms
of collapse equation (\ref{60}) as
\begin{eqnarray*}
\frac{1}{\mathcal{A}_0\mathcal{B}_0}&=&1,~
\frac{\mathcal{A}'_0}{\mathcal{A}_0}=-\frac{m'_o}{r}+\frac{m_o}{r},~\frac{\mathcal{B}'_0}{\mathcal{B}_0}=+\frac{m'_o}{r}-\frac{m_o}{r},~\mathcal{C}_0=r,~
\frac{\mathcal{C}'_0}{\mathcal{C}_0}=\frac{1}{r},\\
\frac{A_0}{B_0^2}&=&1-\frac{3m_o}{r},~\frac{B_0^2}{A_0}=1+\frac{3m_o}{r},~\frac{1}{\mathcal{A}_0}=1+\frac{m_o}{r},~\frac{1}{\mathcal{B}_0}=1-\frac{m_o}{r}.
\end{eqnarray*}
Using these values in collapse equation, we obtain an equation
representing hydrodynamical state for post-Newtonian order like
Newtonian order. The cylindrical system becomes unstable in
post-Newtonian orders if the adiabatic index satisfies the following
inequality
\begin{equation}\label{65}
\Gamma<\frac{\textmd{J}_{\textmd{1pN}}}{\textmd{J}_{\textmd{2pN}}},
\end{equation}
where ${\textmd{J}_{\textmd{1pN}}}$ and
${\textmd{J}_{\textmd{2pN}}}$ are the expressions from
Eq.(\ref{23+}) under post-Newtonian constraints. Following the
Newtonian order cases, we can construct the possibilities of
hydrodynamical equilibrium as well as instable states.

\section{Conclusion}

The collapse process happens when due to some internal or external
disturbance, the matter of object become unbalanced and to maintain
its equilibrium, it collapse down and leads to different structure
i.e., white dwarfs, black holes and stellar groups. Dynamical
instability ranges are used for spherically symmetric object such as
galactic halos, globular clusters etc, while cylindrical symmetry
and plates are associated with the post-shocked clouds at stellar
scale. To study the unstable behavior of spherically symmetric
collapsing object due to its own gravity, the adiabatic index is
used whose numerical range is less then $ \frac{4}{3} $ in GR.
Chandrasekhar \cite{15} gave the direction to study and explore the
dynamical instability in demonstrating the development and shaping
of stellar objects that must be stable against fluctuations. During
these processes, the self-gravitating fluid happens to under go many
phases of dynamical activities that remain in hydrostatic
equilibrium for a short span.

After this equilibrium state, the system is changed from its initial
static phase to perturbed and oscillating phase. It is necessary to
study the evolution of the system immediately after its departure
from the equilibrium state. In this work, we have analyzed the
instability ranges of self-gravitating cylindrically symmetric
collapsing object in $f(T)$ gravity. We have taken isotropic matter
distribution for the interior metric and have found the important
results for interior and exterior regimes. For this purpose, we have
calculated physical quantities like torsion tensor and
super-potential tensor. The selection of tetrad field is crucial for
the framework of $f(T)$ gravity. We have taken non-diagonal tetrad
for the interior spacetime. Torsion scalar has been formulated with
the help of tetrad field, torsion and super-potential tensor in
$f(T)$ gravity.

We have calculated $f(T)$ field equations along with dynamical
equations by using Bianchi identities. A power-law model upto
quadratic torsion scalar term has been taken to examine the dynamics
of collapsing object. Harrison-Wheeler Eq.(\ref{30}) makes
relationship between energy-density and pressure and the adiabatic
index is used to analyze the dynamical instability ranges of the
collapsing object in $f(T)$ gravity. It can be noticed that
Newtonian and post-Newtonian conditions, defined in Eq.(\ref{63})
and Eq.(\ref{65}), are used in collapse equations to find the
corresponding instability ranges. It is also mention that matter
inside inside the cylinder must satisfy some energy conditions. As
discussed in \cite{ec1} that the obtained solutions satisfied some
energy conditions.

\end{document}